\documentclass[pra,twocolumn,showpacs,amsmath,amssymb]{revtex4}
\usepackage{graphicx} 
\usepackage{amsmath}
\usepackage{bm}
\usepackage{hyperref}
\usepackage{dcolumn}

\begin{document}
\title{Calculation of parity nonconserving amplitude and other properties of Ra$^+$}

\author{Rupsi Pal, Dansha Jiang and M. S. Safronova}
\affiliation{Department of Physics and Astronomy, University of
Delaware, Newark, DE 19716-2570, USA}

\author{ U. I. Safronova}
\affiliation{University of Nevada, Reno, NV 89557-0042, USA}

\begin{abstract}
We have calculated parity nonconserving $7s - 6d_{3/2}$ amplitude
$E_{\rm PNC}$ in $^223$Ra$^+$ using high-precision  relativistic all-order
method where all single and double excitations of the
Dirac-Fock wave functions are included to all orders of
perturbation theory. Detailed study of the uncertainty of the parity
nonconserving (PNC)  amplitude is carried out; additional
calculations are performed to estimate some of the missing
correlation corrections. A systematic study of the parity conserving
atomic properties, including the calculation of the energies,
transition matrix elements, lifetimes, hyperfine constants,
quadrupole moments of the $6d$ states, as well as dipole and
quadrupole ground state polarizabilities, is carried out. The
results are compared with other theoretical calculations and
available experimental values.

\end{abstract}


\maketitle
\section{Introduction}

There are two separate reasons for parity violation studies in an
atom: to search for new physics beyond the standard model of the
electroweak interaction by precise evaluation of the weak charge
$Q_w$, and to probe parity violation in the nucleus by evaluating
the nuclear anapole moment. The atomic-physics tests of the standard
model that are completed to date were carried out by comparing
experimental weak charges of atoms $Q_W$, which depend on input from
atomic theory, with predictions from the standard model
\cite{PDBook}. The most precise experimental study to date, a 0.35\%
measurement in Cs was carried out  by the Boulder group \cite{wood1}
using a Stark interference scheme for measuring the ratio of the PNC
 amplitude $E_{\text{PNC}}$ and the vector part of the
Stark-induced amplitude $\beta$ for transitions between states of
the same nominal parity. The value of the  weak  charge in Cs
was ultimately found to be consistent with the theories of the
standard model.
 However, such comparisons provide important
constraints on its possible extensions. A recent
analysis \cite{YCT:07} of parity-violating electron-nucleus
scattering measurements combined with atomic PNC measurements
placed tight constraints on the weak neutral-current lepton-quark
interactions at low energy, improving the lower bound on the scale
of relevant new physics to $\sim$1~TeV.

Experimental measurements of the spin-dependent contribution to the
PNC $6s\rightarrow 7s$ transition in $^{133}$Cs led to a value of
the cesium anapole moment that is accurate to about 14\%
\cite{wood1}. The analysis of this experiment, which required a
calculation of the nuclear spin-dependent PNC amplitude, led to
constraints on weak nucleon-nucleon coupling constants that are
inconsistent with constraints from deep inelastic scattering and
other nuclear experiments, as pointed out in \cite{HAX:01}.
Therefore, new experiments (and associated theoretical analysis) are
needed to resolve the issue. Currently, a microwave experiment to
measure the spin-dependent PNC amplitude in the $7s$ state of Fr
\cite{SB:17} and an isotopic chain experiment in Yb \cite{BER:02}
are underway. We note that  when an experimental study is conducted
in a single isotope, both theoretical and experimental
determinations of  PNC amplitudes are required while the experiments
conducted with isotopic chains should allow to remove the dependence
on the theory. However, accurate theoretical values for a number of
atomic properties are useful for this type of experiments as well.

The present work is motivated by the project that was recently started  at the
Accelerator Institute (KVI) of the University of Groningen
\cite{giri1} to measure PNC amplitude in a single trapped radium
ion. Ra$^+$ is a particularly good candidate for the PNC study owing to
high value of the nuclear charge Z and, correspondingly, large 
expected PNC effects. The $7s-6d_{3/2}$ transition in Ra$^+$ is of
special interest owing to the long life of the $6d_{3/2}$ state and
its sensitivity to both spin independent PNC and spin dependent PNC
~\cite{geetha1}. The $7s-6d_{3/2}$ transition in Ra$^+$ is also
being considered for the development of optical frequency standards
at the same laboratory \cite{sahoo2}. The parity violation
experiments are also accompanied by a number of measurements of
parity-conserving quantities; as a result we have included a
systematical study of such properties in this work.

In summary, we have calculated the PNC amplitude for the
$7s-6d_{3/2}$ transition in $^{223}$Ra$^+$ together with the lifetimes
of the $7p$ and $6d$ states, energy levels for $ns$, $np$, $nd$, and
$nf$ states, transition matrix elements for a number of the E1 and
E2 transitions,  quadrupole moments of the $6d$ states, ground state
dipole and quadrupole polarizabilities, and magnetic-dipole
hyperfine constants A for the $7s, 7p$, and $6d$ states using the
relativistic all-order method. The all-order
 method has proved to be very reliable for calculating
 the properties of alkali-metal atoms and singly ionized monovalent
 ions
 (see, for example, Refs.~\cite{blundell1,safrono1,safrono2,safrono3,safrono4,derevianko1,advances}).
 The effect of Breit interaction on the PNC
   amplitude is also evaluated. The sensitivity of the PNC
   amplitude to the nuclear radius and varying neutron distribution has been studied. Our results
   are compared with other theoretical values and available
   experimental data.

\section{Theory}
\label{method}
 In this section, we briefly discuss the all-order
method which has been used to calculate the wave functions and the
matrix elements necessary to evaluate the observed properties. The
all-order method relies on including all single and double excitations
of the core and valence electrons from the lowest-order
wave function:
\begin{eqnarray}
\label{eq1}
 |\Psi_v \rangle = [1+ \sum_{ma} \rho_{ma} a^{\dag}_m
a_a +\frac{1}{2}\sum_{mnab}\rho_{mnab} a^{\dag}_m a^{\dag}_n a_b a_a \nonumber\\
+\sum_{m\neq v} \rho_{mv} a^{\dag}_m a_v+\sum_{mna}\rho_{mnva}
a^{\dag}_m a^{\dag}_n a_a a_v]{|\Phi_v\rangle}.
\end{eqnarray}
Here, ${|\Phi_v\rangle}$ is the lowest-order atomic wave function
taken to be the frozen-core DF wave function of a state $v$;
$a_{i}^{\dagger}, a_{j}$ are  single-particle  creation and
annihilation operators, $\rho_{ma}$ and $\rho_{mv}$ are the single
core and valence excitation coefficients, and $\rho_{mnab}$ and
$\rho_{mnva}$ are double core and valence excitation coefficients,
respectively. Indices at the beginning of the alphabet, $a$, $b$,
$\cdots$,  refer to occupied core states, those in the middle of
the alphabet $m$, $n$, $\cdots$, refer to excited states, and
index $v$ designates the  valence orbital.

To derive equations for the excitation coefficients,
 the all-order wave function (\ref{eq1}) is substituted into
 the  many-body Schr\"{o}dinger equation
$H | \Psi_v\rangle=E| \Psi_v\rangle, \label{eq2}$ and terms on the
left- and right-hand sides are matched, based on the number and
type of operators they contain.
  Hamiltonian $H=H_0+V_I$
is taken to be the relativistic \textit{no-pair} Hamiltonian:
\begin{eqnarray}
H_{0}&=&\sum_{i=1}^{N}\varepsilon _{i}:a_{i}^{\dag }a_{i}:,  \nonumber \\
V_I&=&\frac{1}{2}\sum_{ijkl}g_{ijlk}:a_{i}^{\dag }a_{j}^{\dag
}a_{l}a_{k}:,
\end{eqnarray}
where $\varepsilon_{i}$ are the single-particle energies, :\,:
designate normal ordering of the operators with respect to closed
core, and $g_{ijkl}$ are the two-body Coulomb matrix elements. The
all-order equations are solved numerically using a finite basis set
of single-particle wave functions which are linear combinations of
B-splines. We have used 70 basis set B-spline orbitals of order 8
defined on a non-linear grid with 500 points within a spherical
cavity of radius 80 a.u. A large spherical cavity is needed to
accommodate all the valence orbitals required for our calculation.
A sufficiently large number of grid points were enclosed within the nucleus
to accommodate the influence of the nucleus on certain atomic
properties such as parity-violating matrix elements and hyperfine
constants.

\begin{table*}
\caption{\label{tab0}  Contributions to the energies of Ra~II: lowest-order (DF) $E^{(0)}$, single-double
Coulomb all-order correlation energy  $E^\text{{SD}}$, third-order
terms not included in the SD value $E^{(3)}_\text{{extra}}$,
  first-order Breit and second-order Coulomb-Breit corrections $B^{(n)}$,
   and Lamb shift $E_\text{
  LS}$.
The total energies $E^\text{{SD}}_\text{ tot}$ are compared with
experimental energies $E_\text{{expt}}$ \protect\cite{nist-web,nist},
 $\delta E$ = $E^\text{{SD}}_\text{ tot}$ - $E_\text{{expt}}$. 
 Our predicted energy values are listed for the $9p_{1/2}$ and 
 $10p_j$ energy levels in separate rows.
 Units: cm$^{-1}$.}
\begin{ruledtabular}
\begin{tabular}{lrrrrrrrrr}
\multicolumn{1}{c}{$nlj$ } & \multicolumn{1}{c}{$E^{(0)}$} &
\multicolumn{1}{c}{$E^\text{{SD}}$} &
\multicolumn{1}{c}{$E^{(3)}_\text{{extra}}$} &
\multicolumn{1}{c}{$B^{(1)}$} & \multicolumn{1}{c}{$B^{(2)}$} &
\multicolumn{1}{c}{$E_\text{ LS}$}&
\multicolumn{1}{c}{$E^\text{{SD}}_\text{ tot}$} &
\multicolumn{1}{c}{$E_\text{{expt}}$} &
\multicolumn{1}{c}{$\delta E^\text{{SD}}$} \\
\hline
$ 7s_{1/2}$& -75898&  -6692& 1152&   147& -250& 33& -81508& -81842&  334\\
$ 6d_{3/2}$& -62356&  -8042& 1152&   155& -398&  0& -69488& -69758&  270\\
$ 6d_{5/2}$& -61592&  -7034&  926&   114& -360&  0& -67947& -68099&  152\\
$ 7p_{1/2}$& -56878&  -4027&  587&   102& -109&  0& -60326& -60491&  165\\
$ 7p_{3/2}$& -52906&  -3020&  433&    63&  -90&  0& -55519& -55633&  114\\
$ 8s_{1/2}$& -36860&  -1745&  316&    46&  -74&  7& -38311& -38437&  126\\
$ 7d_{3/2}$& -31575&  -1590&  245&    39&  -92&  0& -32973& -33098&  125\\
$ 7d_{5/2}$& -31204&  -1456&  204&    29&  -84&  0& -32509& -32602&   93\\
$ 5f_{5/2}$& -28660&  -4438&  371&    11&  -63&  0& -32780& -32854&   74\\
$ 5f_{7/2}$& -28705&  -4159&  353&     8&  -61&  0& -32564& -32570&    6\\
$ 8p_{1/2}$& -30053&  -1298&  201&    39&  -42&  0& -31152& -31236&   84\\
$ 8p_{3/2}$& -28502&  -1034&  156&    25&  -36&  0& -29391& -29450&   59\\
$ 9s_{1/2}$& -22004&   -741&  136&    21&  -33&  2& -22618& -22677&   59\\
$ 9p_{1/2}$& -18748&   -605&   96&    20&  -21&  0& -19259&       &     \\
           &       &       &     &      &     &   & -19305$^{a}$&&\\
$ 9p_{3/2}$& -17975&   -495&   76&    13&  -18&  0& -18399& -18432&   33\\
$ 8d_{3/2}$& -19451&   -683&  105&    18&  -40&  0& -20051& -20107&   56\\
$ 8d_{5/2}$& -19261&   -634&   90&    13&  -37&  0& -19829& -19868&   39\\
$10s_{1/2}$& -14651&   -388&   72&    11&  -18&  1& -14972& -15004&   32\\
$10p_{1/2}$& -12838&   -335&   53&    11&  -11&  0& -13120&       &     \\
           &       &       &     &      &     &   & -13144$^{a}$&&\\
$10p_{3/2}$& -12397&   -278&   43&     7&  -10&  0& -12635&       &     \\
           &       &       &     &      &     &   & -12653$^{a}$&&\\
$ 9d_{3/2}$& -13226&   -366&   56&    10&  -22&  0& -13548& -13578&   30\\
$ 9d_{5/2}$& -13118&   -342&   49&     7&  -20&  0& -13424& -13447&   23\\
$10d_{3/2}$&  -9587&   -221&   34&     6&  -13&  0&  -9780&       &     \\
$10d_{5/2}$&  -9519&   -207&   30&     4&  -12&  0&  -9704&       &     \\
\end{tabular}
\end{ruledtabular}
$^a$ Our predicted values.
\end{table*}

The resulting single-double (SD) excitation coefficients are used to
calculate matrix elements of various one-body operators represented
in the second quantization as $Z=\sum_{ij}z_{ij}a^{\dag}_i a_j$:
\begin{equation}
Z_{wv}=\frac{\langle{\Psi_w} |Z|
\Psi_v \rangle}{\sqrt{{\langle \Psi_v|\Psi_v \rangle }{\langle \Psi_w|\Psi_w \rangle }}}.
\end{equation}

Substituting the expression for the wave function from 
Eq.(\ref{eq1}) in the above equation and simplifying, we get
\begin{equation}
Z_{wv}=\frac{z_{wv}+Z^{(a)}+\cdots+Z^{(t)}}{\sqrt{(1+N_v)(1+N_w)}},
\label{eqr}
\end{equation}
where $z_{wv}$ is the lowest-order DF matrix element and
$Z^{(a)}, \cdots, Z^{(t)}$ and normalization terms $N_i$ are
linear or quadratic functions of the single and double excitation
coefficients {~\cite{blundell1,blundell2}}.
 The expression in
Eq.~(\ref{eqr}) does not depend on the nature of the operator $Z$,
only on its rank and parity.
 Therefore, all matrix elements calculated in this work (E1, M1, E2, hyperfine, and
 PNC matrix elements) are calculated using the same general code.

\begin{table}
\caption{Comparison of the excitation energies important to the
calculation of the $7s-6d_{3/2}$ PNC amplitude.
 All results are in cm$^{-1}$.
\label{tab-en}}
\begin{ruledtabular}
\begin{tabular}{crrrr}
\multicolumn{1}{c}{Transition}& \multicolumn{1}{c}{Present}&
\multicolumn{1}{c}{Ref. \protect\cite{dzuba1}}&
\multicolumn{1}{c}{Ref. \protect\cite{wansbeek1}}&
\multicolumn{1}{c}{Expt.}\\ \hline
$7s - 7p_{1/2}    $&        21182&  21279&  21509&  21351\\
$7s - 7p_{3/2}    $&        25989&  26226&  26440&  26209\\
$6d_{3/2} - 7p_{1/2}$&  9162    &   9468    &   9734    &   9267\\
$6d_{3/2} - 7p_{3/2}$&  13969&  14415&  14665&  14125\\
\end{tabular}
\end{ruledtabular}
\end{table}

Corrections to the all-order equations from the dominant class of
triple excitation terms are also evaluated where needed by including
the term
 $\frac{1}{6}\rho_{mnrvab} a^{\dag}_m a^{\dag}_n a^{\dag}_r a_v a_b a_a |\Phi_v\rangle$
 into the SD wave function (\ref{eq1}) and considering its effect on the energy and
 single valence excitation coefficient equations perturbatively
 (SDpT approach).
 Other classes of triple and higher excitations are included where
 needed using the scaling procedure  by
multiplying single excitation coefficients $\rho_{mv}$ by the
ratio of the ``experimental'' and corresponding (SD or SDpT)
correlation energies \cite{blundell1}. The ``experimental''
correlation energies are determined as the difference of the total
experimental energy and the DF lowest-order values.
 The calculation of the matrix elements is then repeated with the
modified excitation coefficients. We refer the reader to the review
~\cite{advances} and references therein for the detailed description
of the all-order method and its extensions.
 The various atomic properties calculated using the all-order
method described above are discussed in detail in the following
sections.

\section{Properties of Ra$^+$}
\subsection{Energies}

Results of our  calculations of energies for a number of  Ra$^+$
levels  are summarized in Table~\ref{tab0}.  The first six columns
of Table~\ref{tab0} give the lowest-order DF energies $E^{(0)}$, the
all-order SD energies $E^\text{{SD}}$, the part of the third-order
energies omitted in the SD calculation $E^{(3)}_\text{{extra}}$,
 first-order Breit contribution $B^{(1)}$, second-order
Coulomb-Breit $B^{(2)}$ corrections, and Lamb shift contribution,
$E_{\rm LS}$ (see Ref.~\cite{safrono-fr} for detail). We take the sum of these six
contributions to be our final all-order  results,  $E^{\rm
SD}_{\rm tot}$ listed in the seventh column of Table~\ref{tab0}.

The column labeled $\delta E^{\rm SD}$ in Table~\ref{tab0} gives
differences between our {\it ab initio}
 results and the experimental  values \cite{nist,nist-web}.
The SD results are in good agreement with the experimental values
taking into account very large size of the high-order correlation
corrections. We predict the energies of the $9p_{1/2}$, 
$10p_{1/2}$, and $10p_{3/2}$ levels using our theoretical results and 
differences between our and experimental values for the known $np$
levels. The predicted values are listed in Table~\ref{tab0} and are expected to be 
accurate to a few cm$^{-1}$.

\begin{table}
\caption{\label{tab-dip} Comparison of the present results for the
absolute values of the electric-dipole reduced matrix elements in
Ra~II with other theoretical calculations. All results are in
atomic units. The lowest-order DF values are listed in the column
labeled ``DF'' to illustrate the size of the correlation
correction. Negative sign of the DF value for the
$8p_{1/2}-7s_{1/2}$ transition indicates that the lowest-order
value is of the opposite sign with the final result. }
\begin{ruledtabular}
\begin{tabular}{lrrrrr}
\multicolumn{1}{c}{Transition}& \multicolumn{1}{c}{DF}&
\multicolumn{1}{c}{Present}& \multicolumn{1}{c}{Ref.\cite{dzuba1}}
& \multicolumn{1}{c}{Ref.\cite{sahoo2}} &
\multicolumn{1}{c}{Ref.\cite{wansbeek1}}
\\
\hline
$7p_{1/2}-7s_{1/2}$&       3.877&     3.254&     3.224& 3.28&3.31 \\
$7p_{1/2}-8s_{1/2}$&       2.637&     2.517&     2.534&& \\
$7p_{1/2}-9s_{1/2}$&       0.716&     0.702&     0.708&& \\
$7p_{1/2}-6d_{3/2}$&       4.446&     3.566&     3.550& 3.64&3.68\\
$7p_{1/2}-7d_{3/2}$&       4.527&     4.290&     4.358&& \\
$7p_{1/2}-8d_{3/2}$&       1.584&     1.445&     1.432&& \\[0.4pc]
$7p_{3/2}-7s_{1/2}$&       5.339&     4.511&     4.477& 4.54& 4.58\\
$7p_{3/2}-8s_{1/2}$&       4.810&     4.644&     4.663&& \\
$7p_{3/2}-9s_{1/2}$&       1.078&     1.035&     1.036&& \\
$7p_{3/2}-6d_{3/2}$&       1.881&     1.512&     1.504& 1.54&1.56\\
$7p_{3/2}-7d_{3/2}$&       2.488&     2.384&     2.407&& \\
$7p_{3/2}-8d_{3/2}$&       0.733&     0.652&     0.641&& \\
$7p_{3/2}-6d_{5/2}$&       5.862&     4.823&     4.816& 4.92&\\
$7p_{3/2}-7d_{5/2}$&       7.249&     6.921&     6.995&& \\
$7p_{3/2}-8d_{5/2}$&       2.227&     2.011&     1.954&& \\[0.4pc]
$8p_{1/2}-7s_{1/2}$&      -0.125&     0.047&     0.088& 0.04& \\
$8p_{1/2}-8s_{1/2}$&       7.371&     6.949&     6.959&& \\
$8p_{1/2}-9s_{1/2}$&       5.227&     5.012&     5.035&& \\
$8p_{1/2}-6d_{3/2}$&       0.105&     0.049&     0.013& 0.07&\\
$8p_{1/2}-7d_{3/2}$&       10.21&     9.553&     9.540&& \\
$8p_{1/2}-8d_{3/2}$&       7.184&     7.010&     7.104&& \\[0.4pc]
$8p_{3/2}-7s_{1/2}$&       0.625&     0.395&     0.339& 0.50&\\
$8p_{3/2}-8s_{1/2}$&       9.880&     9.294&     9.320&& \\
$8p_{3/2}-9s_{1/2}$&       9.244&     9.022&     9.036&& \\
$8p_{3/2}-6d_{3/2}$&       0.168&     0.144&     0.127& 0.15&\\
$8p_{3/2}-7d_{3/2}$&       4.331&     4.035&     4.028&& \\
$8p_{3/2}-8d_{3/2}$&       4.047&     4.002&     4.034&& \\
$8p_{3/2}-6d_{5/2}$&       0.462&     0.378&     0.347& 0.40& \\
$8p_{3/2}-7d_{5/2}$&       13.37&     12.55&    12.53&& \\
$8p_{3/2}-8d_{5/2}$&       11.68&     11.49&    11.58&& \\[0.4pc]
\end{tabular}
\end{ruledtabular}
\end{table}

We compare our results for the excitation energies important to the
calculation of the $7s-6d_{3/2}$ PNC amplitude with other
theoretical calculations and experiment \cite{nist} in
Table~\ref{tab-en}. The calculations in both Ref.~\cite{dzuba1} and
Ref.~\cite{wansbeek1} use high-precision all-order methods, but
represent very different approaches. The calculations in
Ref.~\cite{dzuba1} are performed using the correlation potential
method. The results of Ref.~\cite{wansbeek1} are obtained using
coupled-cluster method including single, double, and partial triple
excitations. The results of Ref.~\cite{dzuba1} are in better
agreement with experiment for the $7s-7p$ transitions and the results
from the present work are in better agreement with experiment for
the $6d_{3/2}-7p$ transitions. Large discrepancies of the
coupled-cluster results from Ref.~\cite{wansbeek1} for the $6d-7p$
transitions with experiment are somewhat surprising and may indicate
insufficient number of higher partial wave functions in the basis
set. In our calculations, all partial wave up to $l_{max}=6$ are
explicitly included in all calculations and extrapolation for higher
number of partial waves is carried out for the dominant second-order
correlation energy contribution. 

\subsection{Electric-dipole matrix elements}

We calculate all allowed reduced electric-dipole matrix elements
between $ns$, $np$, and $n_1d$ states where $n = 7 - 10$ and $n_1
= 6 - 10$ using the method described above. The subset of these
matrix elements is compared with the correlation potential
calculations of Ref.~\cite{dzuba1} and coupled-cluster
calculations of Refs.~\cite{sahoo2,wansbeek1} in
Table~\ref{tab-dip}. Absolute values of the reduced matrix
elements in atomic units are listed in the table. All present
values with the exception of the $7p_{1/2}-8s$, $7p_{3/2}-8s$, $8p_{1/2}-7s$ and $8p_{3/2}-7s$ transitions
are \textit{ab initio} SD values. For these four transitions, we
used scaling procedure described above to provide recommended
values as we expect the scaled values to be more accurate based on Cs
``best set'' data Ref.~\cite{csus}. 
The calculations of Ref.~\cite{dzuba1} are carried out using
fitted Brueckner orbitals (i.e. include semi-empirical correction
to the correlation operator) and include core polarization, structure
radiation, and normalization corrections. We note that
Ref.~\cite{dzuba1} quotes radial integrals rather than reduced
matrix elements, so we have multiplied their results by the
appropriate angular factors for the purpose of comparison. The
calculations of the Refs.~\cite{sahoo2,wansbeek1} are carried out
using the coupled-cluster method. 

 We have also listed the
lowest-order DF values in the first column of the table to
illustrate the size of the correlation corrections for various
transitions. Negative sign of the DF value for the
$8p_{1/2}-7s_{1/2}$ transition indicates that the lowest-order
values is of the opposite sign with the final result. The
correlation corrections for the primary $7s-7p$ and $7p-6d$
transitions are quite large,  18-25\%. The correlation corrections
for the remaining strong transitions are generally smaller,
2-10\%. All theoretical values are in good agreement for these
transitions. Our values for $7s-7p$ and $7p-6d$ are in better
agreement with results of Ref.~\cite{dzuba1} than those of
Refs.~\cite{sahoo2,wansbeek1}. The agreement is generally poorer
for the transitions with small values of the matrix elements as
expected owing to very large size of the correlation corrections.
Since different methods omit or include somewhat different classes
of the high-order corrections, discrepancies are expected when
such corrections are  large. The issue of the very
small matrix elements, such as $8p-7s$, is also discussed in
Ref.~\cite{dzuba1}.

\subsection{Polarizabilities}

We calculate the static dipole and quadrupole polarizabilities of
the Ra$^+$ ion in its ground $7s$ state. The static polarizability
is calculated as the sum of three terms representing contributions
from the ionic core $\alpha_{c}$, a small counteracting term to
compensate for the excitations from the core states to the valence
state $\alpha_{vc}$, and valence polarizability $\alpha_{v}$:
\begin{equation}
\alpha = \alpha_c+\alpha_{vc}+\alpha_{v}.
\end{equation}

\subsubsection{Dipole polarizability}
 The  valence polarizability contributes over 90\% of
the total value of the electric-dipole polarizability and is
calculated using sum-over-states approach:
\begin{equation}
\alpha_{v}(E1)=\frac{1}{3}\sum_{n}\left(\frac{|\langle 7s||D||np_{1/2}
\rangle|^2}{E_{np_{1/2}}-E_{7s}} + \frac{|\langle 7s||D||np_{3/2}
\rangle|^2}{E_{np_{3/2}}-E_{7s}}\right) \label{alpha1}.
\end{equation}
The sum over $n$ in Eq.~(\ref{alpha1}) converges extremely
fast. In fact, the first term with $n = 7$ contributes 99.8\% of the
total value. As a result, we calculate the first few terms (with $n
= 7 - 10$) using our all-order matrix elements from
Table~\ref{tab-dip} and experimental energies \cite{nist,nist-web} where
available. The remainder $\alpha_v^{\rm tail}$ is calculated in the DF
approximation without loss of accuracy. The ionic core contribution
$\alpha_c$ and term $\alpha_{vc} $ are calculated in the
random-phase approximation (RPA). The RPA core value is expected to be
accurate to better than 5\% (see Ref.~\cite{tchoukova1} and
references therein). All contributions to the dipole polarizability
are listed in Table~\ref{tab-alpha1}. The contributions from $n = 7
- 10$ are given together as $\alpha^{\rm main}_v$. 

The value of the ground
state Ba$^+$ polarizability calculated by the same approach
~\cite{tchoukova1} is in near perfect agreement with the experiment
\cite{lundeen} (to 0.2\%). Moreover, the theoretical SD $6p$
lifetimes in Ba$^+$ are also in excellent agreement with experimental values
\cite{tchoukova1}. We note that lifetime experiments are conducted
entirely differently from the polarizability measurement of
\cite{lundeen}. There are two differences between the Ba$^+$ and
Ra$^+$ dipole polarizability calculations: increased ionic core
contribution and increased size of the correlation corrections. The
core contribution increases from 8\% in Ba$^+$ to 13\% in Ra$^+$ and the
correlation correction contribution to the $7s-7p$ matrix elements
increases by about 3\% (from 16.6\% to 19.1\% for the $7s-7p_{1/2}$
transition). Neither of these changes is expected to significantly
decrease the accuracy of the Ra$^+$ ground state dipole
polarizability in comparison with the Ba${^+}$ one. Therefore, we
expect our value to be accurate to better than 1\%. Our result is in
agreement with the coupled-cluster calculation of
Ref.~\cite{sahoo2}.

\begin{table}
\caption{Contributions to the ground state dipole polarizability of
Ra$^+$. The contributions from the $(7-10)p$ states are given
separately.  Our result is compared with calculation from
Ref.~\cite{sahoo2}.
 All results are in a.u.
\label{tab-alpha1}}
\begin{ruledtabular}
\begin{tabular}{lr}
\multicolumn{1}{c}{Contribution}& \multicolumn{1}{c}{$\alpha_{E1}$}\\
\hline
   $7p_{1/2} - 7s$  & 36.29  \\
   $7p_{3/2} - 7s$  & 56.79   \\
   $8p_{1/2} - 7s$  &  0.00 \\
   $8p_{3/2} - 7s$  &  0.23 \\
   $(9-10)p - 7s$  &  0.04  \\[0.5pc]
   $\alpha^{\rm main}_v$ &   93.35 \\
   $\alpha_{c}$             &   13.74     \\
   $\alpha^{\rm tail}_v$        &   0.11      \\
   $\alpha_{vc}$            &   -0.98     \\
   Total                    &   106.22    \\
   Theory~\cite{sahoo2}     &   106.12    \\
\end{tabular}
\end{ruledtabular}
\end{table}

\begin{table}
\caption{Contributions to the ground state quadrupole polarizability
and the E2 reduced matrix elements of Ra$^+$ in a.u. The comparison
of our result with other theoretical calculation \cite{sahoo2} is
also presented. \label{tab-alpha2}}
\begin{ruledtabular}
\begin{tabular}{lrr}
\multicolumn{1}{c}{Contribution}& \multicolumn{1}{c}{E2}&
\multicolumn{1}{c}{$\alpha_{E2}$}
\\ \hline
   $6d_{3/2} - 7s$  & 14.74(15) & 789(13) \\
   $6d_{5/2} - 7s$  & 18.86(17) & 1136(16)\\
   $7d_{3/2} - 7s$  & 14.21(30) & 182(3)  \\
   $7d_{5/2} - 7s$  & 16.49(38) & 243(4)  \\
   $8d_{3/2} - 7s$  & 5.63(4)   & 22.6(2) \\
   $8d_{5/2} - 7s$  & 6.79(6)   & 32.6(2)   \\
   $9d_{3/2} - 7s$  & 3.30(3)   & 7.0(1) \\
   $9d_{5/2} - 7s$  & 4.03(3)   & 10.4(1) \\
   $10d_{3/2} - 7s$ & 2.27(3)   & 3.1  \\
   $10d_{5/2} - 7s$ & 2.79(3)   & 4.7   \\[0.5pc]
   $\alpha^{\rm main}_v$ &  &  2430(21)  \\
   $\alpha^{\rm tail}_v$        & &  35(10)     \\
   $\alpha_{c}$             & &  68(12)   \\
   Total                    & &  2533(26)   \\
   Theory~\cite{sahoo2}     & &  2547.5    \\
\end{tabular}
\end{ruledtabular}
\end{table}

\subsubsection{Quadrupole polarizability}

The  valence part of the quadrupole polarizability is calculated
using the sum-over-states approach as:
\begin{equation}
\alpha_{v}(E2)=\frac{1}{5}\sum_{n}\left(\frac{|\langle 7s||Q||nd_{3/2}
\rangle|^2}{E_{nd_{3/2}}-E_{7s}} + \frac{|\langle 7s||Q||nd_{5/2}
\rangle|^2}{E_{nd_{5/2}}-E_{7s}}\right) \label{alpha2}.
\end{equation}

All contributions to the quadrupole polarizability are listed in
Table~\ref{tab-alpha2}. The correlation correction to the E2 matrix
elements is dominated by a single term among twenty terms in the
numerator of Eq.~(\ref{eqr}). As described in detail in
Ref.~\cite{tchoukova1}, additional omitted correlation correction to
this term may be estimated by the scaling procedure described above.
The scaling modifies the SD results by 0.7 to 2.3\%
depending on the transition. We have also carried out the \textit{ab
initio} all-order calculation with inclusion of the triple valence 
excitation coefficients  as described in the Section~\ref{method}
 (SDpT
approach). The scaling procedure was repeated starting from the SDpT
approximation for the dominant $7s - 6d_{3/2}$ and $7s-6d_{5/2}$
transitions. These additional calculations allow us to directly
evaluate the uncertainty in our calculations since they produce
different evaluations of the omitted correlation correction. We take
the uncertainty in the calculation of the $7s - 6d_{3/2}$ and
$7s-6d_{5/2}$ to be the maximum of the difference of out final SD
scaled results with \textit{ab initio} and scaled SDpT data.   We
note that SD approach generally underestimates the correlation
energy and SDpT approach generally overestimates the correlation
energy used in the scaling procedure. The scaled SD and SDpT results
are rather close, further confirming the validity of this procedure
and of our uncertainty estimate. Therefore,  we take the uncertainty of the remaining
transitions to be the difference of the
final SD scaled and \textit{ab initio} SDpT values. The
resulting final matrix elements and their uncertainties are listed
in Table~\ref{tab-alpha2} in column labeled ``E2''. The relative
uncertainty of the corresponding polarizability values is
twice the relative uncertainty of the matrix elements since we
assume the experimental energies be accurate to all figures quoted.
The sum over $n$ converges far slower than in the case of the dipole
polarizability so calculating a first few terms to high precision is
essential to obtain an accurate final value. The tail contribution,
while small, is significant and has to be treated with care. We
estimated that DF value for the main ($n = 6 - 10$) term is larger
than our final all-order result by 22\%. Therefore, we decrease the
DF tail of 45~a.u. by 22\% and take the difference of the DF tail and
the final adjusted value to be its uncertainty. The core
contribution is calculated in the RPA approximation; we take the
difference between DF and RPA core values to be the uncertainty of the core contribution.
Our final value is in agreement with the result of
Ref.~\cite{sahoo2}.

\subsection{Lifetimes of the $7p$ and $6d$ states}

The lifetimes $\tau$ of the $7p$ and $6d$ states in Ra$^+$ are
calculated as the inverse of the sum of the transition probabilities
$A$. The $7p$ states decay via strong electric-dipole transitions.
Total of five E1 transitions contribute to the lifetimes of these
two states: $7p_{1/2}-7s$, $7p_{1/2}-6d_{3/2}$, $7p_{3/2}-7s$,
$7p_{3/2}-6d_{3/2}$, and  $7p_{3/2}-6d_{5/2}$. The electric-dipole
transition rates are calculated using formula

\begin{equation}
A^{E1}_{if} = \frac{2.02613 \times 10^{18}}{\lambda^3}
 \frac{|{\langle i||D||f \rangle }|^2}{2j_i+1} s^{-1},
 \end{equation}
where $\lambda$ is the wavelength of the transition in \AA~and
$\langle i||D||f \rangle$ is the electric-dipole reduced matrix element
in atomic units. We use the experimental wavelength \cite{nist,nist-web} and
our all-order matrix elements listed in Table~\ref{tab-dip} when
evaluating the transition rates. The results are summarized in
Table~\ref{tab7p}. We find that while the contributions of the $7s -
7p$ transitions to the $7p$ lifetimes are dominant, the contributions of the $7p - 6d$ transitions are
significant (over 10\%).

\begin{table}
\caption{Contributions to the lifetimes of the $7p_{1/2}$ and $7p_{3/2}$
states. The transitions rates $A$ are given in $10^6~$s$^{-1}$ and
the lifetimes are given in ns. \label{tab7p} }
\begin{ruledtabular}
\begin{tabular}{lrclr}
 \multicolumn{2}{c}{$7p_{1/2}$}     &
 \multicolumn{1}{c}{}     &
\multicolumn{2}{c}{$7p_{3/2}$}      \\ \hline
   $A(7p_{1/2} - 7s)$       & 104.4  && $A(7p_{3/2} - 7s)$      &   185.5       \\
   $A(7p_{1/2} - 6d_{3/2})$ & 10.3   && $A(7p_{3/2} - 6d_{3/2})$&     3.3    \\
   $\sum A        $          & 114.7 && $A(7p_{3/2} - 6d_{5/2})$&22.8 \\
   $\tau(7p_{1/2}) $         & 8.72~ns      &&    $\sum A        $  &211.6         \\
                             &        &&    $\tau(7p_{3/2})$  & 4.73~ns       \\
   \end{tabular}
\end{ruledtabular}
\end{table}

Only one transition, $6d_{3/2} - 7s$, has to be considered for the
calculation of the $6d_{3/2}$ lifetime. The corresponding transition
rate is calculated as
\begin{equation}
A^{E2}_{if} = \frac{1.11995 \times 10^{18}}{\lambda^5}
\frac{|{\langle i||Q||f \rangle }|^2}{2j_i+1} s^{-1},
\end{equation}
where $\lambda$ is the wavelength of the transition in \AA~and
$\langle i||Q||f \rangle$ is the electric-quadrupole reduced matrix
element in atomic units.

Two transitions have to be considered in the calculation of the
$6d_{5/2}$ lifetime: E2 $6d_{5/2} - 7s$ transition and M1 $6d_{5/2}
- 6d_{3/2}$ transition. The M1 transition rate is calculated as
\begin{equation}
A^{M1}_{if} = \frac{2.69735 \times 10^{13}}{\lambda^3}
\frac{|{\langle i||M1||f \rangle }|^2}{2j_i+1} s^{-1}.
\end{equation}
We use the experimental wavelengths \cite{nist,nist-web} and our all-order
matrix elements listed in Table~\ref{tab-alpha2} when evaluating the
E2 transition rates. Our result for the  reduced M1
 $6d_{5/2} - 6d_{3/2}$ matrix element is 1.55~a.u. The E2 and M1 transition rates
contributing to the  $6d_{5/2}$ lifetime are 3.255~s$^{-1}$ and
0.049~s$^{-1}$. We verified that the contribution of the $6d_{5/2} -
6d_{3/2}$ E2 transition is negligible.

Our results for the $6d_{3/2}$ and $6d_{5/2}$ lifetimes  are
presented in Table \ref{tab6d} together with other theoretical
values. Our values for the lifetimes of the $6d$ states are in
better agreement with those published by Dzuba \textit{et al.}
\cite{dzuba1} than with the results of Sahoo \textit{et al.}
\cite{sahoo2}; however, the discrepancies with 
Ref.~\cite{sahoo2} are small. We also list the uncertainties of our
values in Table~\ref{tab6d}. The relative uncertainties in our
values of the $6d$ lifetimes are twice the relative uncertainties in
the values of the E2 matrix elements listed in
Table~\ref{tab-alpha2}. We note that the estimated uncertainties
quoted in Ref.~\cite{sahoo2} are obtained by carrying out
calculations with different bases; i.e. they are numerical
uncertainties resulting from the particular choice of the basis set
and do not include estimation of the missing correlation effects. In
our calculations, the basis set is complete (70 splines for each
partial wave) and increasing its size does not change the result.
Our uncertainties include  estimation of the terms beyond triple
contributions as described above as well as uncertainty owing to
truncation of the partial waves above $l>6$. Therefore, while our
uncertainty is higher for $6d_{3/2}$ state than the one quoted in Ref.~\cite{sahoo2}, it represents an attempt
to provide an actual boundary for the recommended value of this
lifetime.
\begin{table}
\caption{Lifetimes of the $6d_{3/2}$ and $6d_{5/2}$ states of Ra$^+$
in seconds. Comparison of our results with other theoretical
calculations is presented.\label{tab6d} }
\begin{ruledtabular}
\begin{tabular}{lllll} \multicolumn{1}{c}{Term}&
\multicolumn{1}{c}{$\tau(6d_{3/2})$}&
\multicolumn{1}{c}{$\tau(6d_{5/2})$}\\
\hline
   Present                & 0.638(10) &  0.303(4) \\
   Theory~\cite{sahoo2}   & 0.627(4)   &  0.297(4) \\
   Theory~\cite{dzuba1}  &  0.641      &  0.302 \\
   \end{tabular}
\end{ruledtabular}
\end{table}
\subsection{Quadrupole moments of the $6d$ states }

We also calculated the values of the quadrupole moments of the $6d_{3/2}$
and $6d_{5/2}$ states since these properties are of interest to the
investigation of possible use of Ra$^+$ for the development of
optical frequency standard \cite{sahoo2}. The quadrupole moment $\Theta(\gamma J)$ can be expressed 
via the reduced matrix element of the quadrupole operator $Q$ as
\begin{equation}
\Theta(\gamma J)=\frac{(2J)!}{\sqrt{(2J-2)!(2J+3)!}}
 \left<\Psi( \gamma J) \left\|  Q \right\|
 \Psi( \gamma J )\right>.
\end{equation}
The calculation follows that of the E2 matrix elements. As
in the case of the E2 $7s - nd$ matrix elements, a single
correlation correction term is dominant, and the omitted correlation
contributions may be estimated via the scaling procedure. We have
conducted four different calculations: \textit{ab initio} SD and
SDpT, and scaled SD and SDpT ones to evaluate the uncertainty in the
final values. The results are summarized in Table~\ref{quad}. The
correlation correction to the quadrupole moments is on the order of
20\%. Our values are compared with coupled-cluster calculation of
Ref.~\cite{sahoo2}. Our results are lower than that of
Ref.~\cite{sahoo2}. This issue has been discussed in detail in
Ref.~\cite{quad-new} where we have demonstrated that CCSD(T) method
may overestimate quadrupole moments by a few percent owing to the
cancellation of various terms. Omission of orbitals with $l>4$ from
the basis set may also lead to higher values.

\begin{table}
\caption{Quadrupole moments of the $6d_{3/2}$ and $6d_{5/2}$ states in
Ra$^+$ in a.u. \label{quad} }
\begin{ruledtabular}
\begin{tabular}{lcccccc}
 \multicolumn{1}{c}{State}&
\multicolumn{1}{c}{SD}& \multicolumn{1}{c}{SDpT}&
\multicolumn{1}{c}{SD$_{sc}$}& \multicolumn{1}{c}{SDpT$_{sc}$}&
\multicolumn{1}{c}{Final}&
\multicolumn{1}{c}{Ref.~\cite{sahoo2}}\\
 \hline
   $6d_{3/2}$ &  2.814 & 2.868 & 2.839 & 2.829  & 2.84(3) & 2.90(2)\\
   $6d_{5/2}$ &  4.311 & 4.380 & 4.342 & 4.329  & 4.34(4) & 4.45(9)\\
   \end{tabular}
\end{ruledtabular}
\end{table}

\subsection{Magnetic-dipole hyperfine constants}

\begin{table}
\caption{Magnetic-dipole hyperfine  constants A(MHz) for
the $7s$, $7p_{1/2}$, $7p_{3/2}$, $6d_{3/2}$, and $6d_{5/2}$ states in $^{223}$Ra$^+$ calculated using 
SD and SDpT all-order approaches. Lowest-order (DF) values are also listed to illustrate the size of the correlation corrections. 
The present values are compared with other theoretical~\cite{wansbeek1,sahoo2} and experimental values from Refs.~\cite{exp_hyp1,exp_hyp2}.
\label{hyp} }
\begin{ruledtabular}
\begin{tabular}{lrrrrrr}
\multicolumn{1}{c}{State}&
 \multicolumn{1}{c}{DF} &
\multicolumn{1}{c}{SD}& 
\multicolumn{1}{c}{SDpT}&
\multicolumn{1}{c}{Ref.~\cite{wansbeek1}}&
\multicolumn{1}{c}{Ref.~\cite{sahoo2}}&
\multicolumn{1}{c}{Expt.}
\\ \hline
$7s$	    &  2614	&  3577&	 3450	 &3557  &3567 & 3404(2)\\
$6d_{3/2}$&	52.92	& 81.51&	 79.56 &79.80 &77.08	     &	 \\
$6d_{5/2}$&	19.24	&-23.98&	-24.08 &	    &	-23.90     & \\
$7p_{1/2}$&	444.5	& 699.5&	671.5	 &671.0 &666.9     	& 667(2)       \\
$7p_{3/2}$&	33.91	& 56.62&	54.40	 &56.53 &56.75	    & 56.5(8)       \\
\end{tabular}
\end{ruledtabular}
\end{table}

Our results for the magnetic-dipole hyperfine constants $A$(MHz) in
$^{223}$Ra$^+$ are compared with theory
\cite{sahoo2,wansbeek1} and experiment \cite{exp_hyp1,exp_hyp2} in Table~\ref{hyp}. The gyromagnetic ratio
$g_I$ for $^{223}$Ra is taken to be $g_I=0.1803$ and corresponds to the value
 $\mu_I=0.2705(19)~\mu_N$ from Ref.~\cite{mu}. We note that the magnetic moment of 
 $^{223}$Ra have not been directly measured but recalculated from 
 measurements of $^{213}$Ra and $^{225}$Ra nuclear magnetic moments in Ref.~\cite{mu}.
   The magnetization distribution is modeled by a Fermi distribution with
the same parameters as our charge distribution ($c$=
6.862~fm and 10\%-90\% thickness parameter is taken
to be $t$=2.3~fm). The lowest-order values are also listed to
demonstrate the size of the correlation corrections for various
states. The triple contributions are important for the hyperfine
constants and are partially included as described in
Section~\ref{method}. These values are listed in column labeled
``SDpT''. The SD values are also listed for comparison in column
labeled ``SD''.

 The value $g_I=0.18067$ that corresponds to the 
rounded off value $\mu_I=0.271(2)~\mu_N$ from \cite{mu} was used in Ref.~\cite{sahoo2}. 
The values for $A/g_I$ were quoted in Ref.~\cite{wansbeek1}, so we multiplied their values by 0.18067 
for comparison. The differences between our results and experimental values are 1.3\%, 0.7\%, and 4\% for 
$7s$, $7p_{1/2}$, and $7p_{3/2}$ states, respectively. We note that the uncertainty in the value of the 
nuclear magnetic moment is 0.7\%.  Larger difference of the  $A(7p_{3/2})$ SDpT value  with the experiment 
is similar to that one in Cs \cite{safrono1}, where the difference of the SDpT value for  the $6p_{3/2}$
magnetic-dipole hyperfine constant with experiment
is 3.5\%.  Interestingly, the Cs SDpT values are below the experimental ones while the 
Ra$^+$ SDpT results are above the experimental values. This can be explained by the uncertainty in the
treatment of the finite size correction, uncertainty in the value of Ra nuclear magnetic moment, and 
the difference in the size and distribution of the correlation corrections in Cs and Ra$^+$.

\section{Parity nonconservation}

\begin{table*}
\caption{Contributions to the $E^{\rm main}_{\rm PNC}$ in $^{223}$Ra$^+$ in
units of $10^{-11}i|e|a_0(-Q_W/N)$. $D$ and $H_{\rm PNC}$ are dipole and 
PNC matrix elements, respectively. Reduced electric-dipole matrix elements are
listed for consistency with previous tables; they need to be multiplied by $1/\sqrt{6}$
to obtain relevant values of $\langle i|D|j\rangle$ ($m_j = 1/2$ for all states).
All values are in a.u. Our results are compared with calculations of Ref.~\cite{dzuba1}. \label{pnc1}  }
\begin{ruledtabular}
\begin{tabular}{cccccc} 
    $n$&$\langle 6d_{3/2}\|D\|np_{1/2}\rangle$&$\langle np_{1/2}|H_{\rm PNC}|7s\rangle$&$E_{7s}-E_{np_{1/2}}$&$E_{\rm PNC}$&Ref.~\cite{dzuba1}\\
    \hline
        $7$   & 3.566 & -2.665 & -0.0973 & 39.882& 40.69\\
        $8$   & 0.049 & -1.590 & -0.2306 & 0.137 &0.11\\
        $9$   & 0.017 & -1.124 & -0.2849 & 0.027 &0.02\\
        $10$  & 0.008 & -0.841 & -0.3130 & 0.009 &\\
      \hline       
 $n$&$\langle 6d_{3/2}|H_{\rm PNC}|np_{3/2}\rangle$ & $\langle np_{3/2}\|D\|7s\rangle$ &$E_{6d_{3/2}}-E_{np_{3/2}}$&$E_{\rm PNC} $&Ref.~\cite{dzuba1}\\
\hline
    $7$ & -0.047 &  -4.551 &  -0.0644 & -1.348& -2.33\\
    $8$ & -0.040 &  -0.405 &  -0.1837 & -0.036& -0.05\\
    $9$ & -0.032 &  -0.140 &  -0.2339 & -0.008&-0.01\\
    $10$& -0.026 &  -0.069 &  -0.2602 & -0.003& \\
\end{tabular}
\end{ruledtabular}
\end{table*}

Nuclear-spin independent PNC effects in atoms are caused by the  exchange of a virtual $Z_0$
 boson between an electron of the atom and a quark in the nucleus, or between
 two atomic electrons~\cite{bookpnc}. The second effect is extremely small and will not
 be considered in this work. The dominant PNC interaction
 between an atomic electron and the nucleus is described by a Hamiltonian
 $A_eV_N$, which is the product of axial-vector electron current
 $A_e$ and vector nucleon current $V_N$. 
 The PNC interaction leads to a non-zero amplitude for transitions otherwise forbidden by the
  parity selection rule, such as the $6d_{3/2}-7s$ transition in singly ionized radium.
 Combining experimental measurements and theoretical calculations
 of the PNC amplitude permits one to infer the value of the weak charge $Q_W$
 for precise atomic-physics tests of the standard model.

The $7s-6d_{3/2}$ PNC amplitude in Ra$^+$ can be evaluated as a sum over states:
\begin{eqnarray}
E_{\rm PNC}&=&\sum_{n=2}^{\infty}
\frac{\langle 6d_{3/2}|D|np_{1/2}\rangle
\langle np_{1/2}|H_{\rm PNC}|7s\rangle}
 {E_{7s}-E_{np_{1/2}}} \nonumber \\
&+&\sum_{n=2}^{\infty}\frac{
\langle 6d_{3/2}|H_{\rm PNC}|np_{3/2}\rangle\langle np_{3/2}|D|7s\rangle}{E_{6d_{3/2}}-E_{np_{3/2}}},
\label{eq_pnc}
\end{eqnarray}
where $D$ is the dipole transition operator. The values of $m_j$ are customary taken to be $m_j=1/2$ for all states.
 The PNC Hamiltonian $H_{\rm PNC}$ is given by
\begin{equation}
        H_{\rm PNC}= \frac{G_F}{2\sqrt{2}}Q_W \gamma_5 \rho(r), \label{h1}
\end{equation}
where $G_F$ is the universal Fermi coupling constant, $Q_W$ is
the weak charge 
and $\gamma_5$ is the Dirac matrix associated with pseudoscalars. The quantity $\rho(r)$ is
a nuclear density function, which is approximately the neutron density. In our calculations,
we model $\rho(r)$ by the charge form factor, which is taken to be a Fermi distribution with
50\% radius $c_{\rm PNC}=c_{\rm charge}=$6.8617~fm \cite{radius} and 10\%-90\% thickness parameter $t=2.3$~fm for $^{223}$Ra$^+$ ,
i.e. we take $\rho(r)$ to be the same distribution as the charge distribution used our entire all-order
calculation of the Ra$^+$ wave functions and corresponding properties. We also
 investigate how the PNC amplitude vary with changes in both  $c_{\rm PNC}$ and $c_{\rm charge}$. 

The sum over $n$ in Eq.~(\ref{eq_pnc}) converges very fast in our case, and only first few
terms need to be calculated accurately. Therefore, 
we divide our calculation of $E_{\rm PNC}$  into three parts:
a main term $E_{\rm PNC}^{\rm main}$ that 
consists of the sum over states with $n = 7 - 10$,
a tail $E_{\rm PNC}^{\rm tail}$ which is the sum over states
with $n=11,\dots,\infty$, and the contribution $E_{\rm PNC}^{\rm auto}$ 
from autoionizing states given by the terms with $n = 2 - 6$.
The calculation of the main term is illustrated in Table~\ref{pnc1}, where we list the ``best set'' of the 
dipole and PNC matrix elements used in our calculation as well as relevant energy differences. 
  The final electric-dipole matrix elements are taken to be \textit{ab initio} single-double all-order results (following the 
  comparison of the similar Cs and Ba$^+$ results with experiment \cite{safrono1,tchoukova1}).
  Reduced electric-dipole matrix elements are
listed for consistency with previous tables; they need to be multiplied by $1/\sqrt(6)$
to obtain relevant values of $\langle i|D|j\rangle$ ($m_j = 1/2$ for all states).
  The final PNC matrix elements for the $6d_{3/2} - 7p_{1/2}$ and $6d_{3/2} - 8p_{1/2}$
transitions are taken to be SD all-order scaled values since the contribution that can 
be accounted for by scaling is the dominant one for these cases; remaining PNC
matrix elements are taken to be \textit{ab initio} SD values. Experimental energies
are used where they are available, our predicted energy values from Table~\ref{tab0}
are used for the $9p_{1/2}$, $10p_{1/2}$, and $10p_{3/2}$
 levels. Our results are compared with results of Ref.~\cite{dzuba1} calculated using the correlation potential method. 
The main part of the PNC amplitude is overwhelmingly dominated by a single term listed in the first row of Table~\ref{pnc1}.
Our result for this term slightly differs from the calculation in Ref.~\cite{dzuba1} (by 2.2\% ). However, the Ref.~\cite{dzuba1}
does not list the Ra$^+$ isotope for which the calculation has been conducted. Since the value of the PNC amplitude is multiplied by the 
neutron number in the present commonly accepted units of $10^{-11}i|e|a_0(-Q_W/N)$,
 the difference between values for the PNC amplitudes for $^{223}$Ra$^+$ and   $^{226}$Ra$^+$ is 2\% just owing to 
 $138/135$ neutron number ratio. Therefore, the difference may be either explained by the simple
 isotope rescaling, difference in the choice of the nuclear density function parameters, or differences in the 
 treatment of the correlation correction. The only significant discrepancy between our calculation and that of 
 Ref.~\cite{dzuba1} is in the other term with $n=7$ (-1.35 vs. -2.33). This difference has to result from the 
 differences in the treatment of the correlation correction since this entire value comes from the correlation effects. 
 Taking into account that the DF value for this term is consistent with zero and random-phase approximation (RPA)
 result, -4.08, is larger than the all-order value by nearly a factor of 3, such discrepancy is not very surprising.

 To provide some estimate of the uncertainty in the calculation of the main term, we conduct the 
 ``scatter'' analysis of the data following the calculation of the Cs PNC amplitude \cite{prd}.  In such
 analysis, sets of data for dipole matrix elements, PNC matrix elements, and energies are
 varied (i.e., taken to be SD, SDpT, expt.) and the scatter in the final PNC values is analyzed. 
 The results are summarized in  Table~\ref{pnc2}. Our final value (corresponding to data in Table~\ref{pnc1}) is listed in the last row of Table~\ref{pnc2}.
  We note that essentially entire difference in the 
  results comes from the dominant term (first row of Table~\ref{pnc1}), and the variation in all other terms is 
  insignificant.  Therefore, the possible uncertainty in the next term (-1.35), which is bound to be 
  substantial, can not be evaluated by this approach.  While we have included the values with SDpT 
 dipole matrix elements, there is  no reason to expect these data to be more accurate than SD values.
 This conclusion is based on the 
 breakdown of the correlation correction contributions and comparisons of the similar calculations in other monovalent systems
 that demonstrate cancellation of some missing effects in SD approximation but not in SDpT one. As a result, we
 conclude that the uncertainty in the dominant term owing to the Coulomb correlation correction is probably 
 on the order of 2\%. We note that completely  \textit{ab initio} SD value is in good agreement with 
 our final value.  Measurement of the $6d_{3/2} - 7p_{1/2}$ oscillator strength would help to 
 reduce this uncertainty.

\begin{table}
\caption{`Scatter'' analysis of the main part of the PNC amplitude ($n = 7 - 10$) in $^{223}$Ra$^+$. 
Lowest-order DF and random-phase  RPA values are listed for reference.  
SD labels single-double all-order values, SDpT values include partial triple contributions.  \label{pnc2} }
\begin{ruledtabular}
\begin{tabular}{cccc} \multicolumn{1}{c}{Energies}&
\multicolumn{1}{c}{$\langle i|D|j \rangle$\vspace{0.1cm}}&
\multicolumn{1}{c}{$\langle i|H_{PNC}|j \rangle$\vspace{0.1cm}}&
\multicolumn{1}{c}{$E^{\rm main}_{\rm PNC}$}
\\ \hline
 DF & DF &DF & 38.95\\
 DF & RPA & RPA & 37.10\\
     SD & SD & SD & 39.05\\
     Expt. & SD & SD & 39.65 \\
     Expt. & SDpT & SD & 40.22\\
     Expt. & SD & SDpT & 38.09\\
     Expt. & SDpT & SDpT & 38.65\\
        Expt. & SD &SD$_{sc}^{a}$ & 38.66\\
\end{tabular}
\end{ruledtabular}
$^{a}$ Scaled values are used for the $7p_{1/2}-7s$ and $8p_{1/2}-7s$ matrix elements only, remaining data are taken to be SD. 
\end{table}

We calculate remaining terms $E_{\rm PNC}^{\rm tail}$ and  $E_{\rm PNC}^{\rm auto}$ in both
DF and RPA approximations. The RPA results are listed in Table~\ref{pnc3} together with 
our total value for the PNC amplitude. The corresponding DF results are $E_{\rm PNC}^{\rm auto}=4.8$ 
and $E_{\rm PNC}^{\rm tail}=1.2$. The correction due to Breit interaction is 
obtained in the DF approximation. Our final value is compared with 
other calculations from Refs.~\cite{wansbeek1,dzuba1}. 
Our result for the   terms with $n<7$ and $n>9$ (6.8) is in reasonably good agreement with the value from Ref.~\cite{dzuba1}
(7.5). The notable feature of Table~\ref{pnc3} is an
excellent agreement of all rather different high-precision calculations despite  relatively large possible uncertainties in 
various terms, with the exception of the mixed-states result~\cite{dzuba1}. 
Further calculations as well as experimental measurements will be necessary to achieve 1\% accuracy 
in the PNC amplitude.

\begin{table}
\caption{Contribution to the $E_{PNC}$ in $^{223}$Ra$^+$ and
comparison with other theory. Our value for $^{226}$Ra$^+$ is obtained by reducing our 223 value by 0.2\% owing to the 
correction for the different nuclear parameters and multiplying by 138/135 
neutron number ratio. All results are in units
of $10^{-11}i|e|a_0(-Q_W/N)$. \label{pnc3} }
\begin{ruledtabular}
\begin{tabular}{llr} \multicolumn{1}{l}{Isotope}& \multicolumn{1}{l}{Term}&
\multicolumn{1}{r}{Value}
\\ \hline
 223 &  $E^{\rm main}_{\rm PNC}$  & 38.66 \\
 223 &   $E_{\rm PNC}^{\rm tail}$ & -0.02 \\ 
 223 &   $E_{\rm PNC}^{\rm auto}$  & 6.83 \\ 
 223 & Breit  & -0.29 \\
 223 &             Total  & 45.18 \\[0.5pc]
  226 &             Total  & 46.09 \\[0.5pc]
  &Mixed states~\cite{dzuba1}  & 42.9 \\
&Sum over states~\cite{dzuba1} & 45.9 \\
  226   &      CCSD~\cite{wansbeek1} & 46.1 \\
  226   &   CCSD(T)~\cite{wansbeek1} & 46.4 \\
\end{tabular}
\end{ruledtabular}
\end{table}

\begin{table}
\caption{Dependence of the lowest-order Ra$^+$ PNC amplitude on the parameters of the nuclear
distributions $c_{\rm charge}$(fm) and $c_{\rm PNC}$(fm). The parameter $c_{\rm charge}$ is used in the charge distribution in the all-order
wave function calculations. The parameter $c_{\rm PNC}$ is used in the modeling
the nuclear density function in the PNC Hamiltonian. The variation of the given 
parameter is listed in \% for convenience. 
The units for the PNC amplitude is $10^{-11}i|e|a_0(-Q_W/N)$.\label{pnc4} }
\begin{ruledtabular}
\begin{tabular}{cccccr}
 \multicolumn{1}{c}{$c_{\rm charge}$}&
\multicolumn{1}{c}{$\delta c_{\rm charge}$}&
\multicolumn{1}{c}{$c_{\rm PNC}$}&
\multicolumn{1}{c}{$\delta c_{PNC}$}&
\multicolumn{1}{c}{$E_{\rm PNC}^{\rm DF}$}&
\multicolumn{1}{c}{$\delta E_{\rm PNC}^{\rm DF}$}
\\ \hline
6.8617& 0\%		&6.8617& 		&44.913&	\\
6.8960& 0.5\%	&6.8617& 		&44.853&	-0.13\%\\
6.9303& 1\%		&6.8617& 		&44.792&	-0.27\%\\
6.9989& 2\%		&6.8617&  		&44.671&	-0.54\%\\
7.2048& 5\%		&6.8617& 		&44.310&	-1.34\%\\
      &       &      &   &      &  \\
6.8617& 		&6.8960&	0.5\%	&44.875&    -0.08\%\\
6.8617& 		&6.9303&	1\%	&44.837&    -0.17\%\\
6.8617& 		&6.9989&	2\%	&44.761&    -0.34\%\\
6.8617& 		&7.2048&	5\%	&44.531&    -0.85\%	\\
      &&&&&\\
6.8960& 0.5\%	&6.8960&	0.5\%  &44.815&    -0.22\%\\
6.9303& 1\%		&6.9303&	1\%	&44.717&    -0.44\%\\
6.9989& 2\%		&6.9989&	2\%	&44.523&    -0.87\%\\
7.2048& 5\%		&7.2048&	5\%	&43.954&    -2.14\%\\
\end{tabular}
\end{ruledtabular}
\end{table}

We also investigated the dependence of the PNC amplitude 
 on the values of the nuclear
distribution parameters $c_{\rm charge}$ and $c_{\rm PNC}$. As we described in the beginning of this section, 
the parameter $c_{\rm charge}$ is used in the charge distribution in the all-order
wave function calculations. The parameter $c_{\rm PNC}$ is used in the modeling
of the nuclear density function $\rho(r)$ in the PNC Hamiltonian given by Eq.~(\ref{h1}). Both are modeled by the Fermi distributions;
the all-order calculation is carried with both half-density parameters being equal to 6.8617~fm \cite{radius}.
Since the DF result is rather close to the final value owing to various cancellations, it is sufficient
to carry out this study using DF data. 
The results are summarized in Table~\ref{pnc4}, where we list $E_{\rm PNC}^{\rm DF}$
calculated with varying values or either one or both parameters. The variation of the given 
parameter is listed in \% for convenience. 
The results show that possible 
uncertainty in the PNC amplitude owing to the uncertainty in the value of the charge radius (that is unlikely to be large) 
is negligible in comparison with the uncertainty in the correlation correction. For example, difference in the rms radii for A=223 and 
A=226 isotopes corresponds to the change in $c_{\rm charge}$ that is on the order of 0.5\% resulting in only 0.2\% change in the PNC 
amplitude. Possible variance in the density $\rho(r)$ in Eq.~(\ref{h1}) which is approximately neutron density is higher, but 
even 5\% change in $c_{\rm PNC}$ with the fixed value of the $c_{\rm charge}$ leads to 0.85\%  change
in the PNC amplitude value. Table~\ref{pnc4} may be used to recalculate the values of the PNC amplitude
between different isotopes since the change in $E_{\rm PNC}$ with the nuclear parameters is essentially linear.

\section{Conclusion}

We have calculated the energies,
transition matrix elements, lifetimes, hyperfine constants,
quadrupole moments of the $6d$ states, as well as dipole and
quadrupole ground state polarizabilities, and PNC amplitude in $^{223}$Ra$^+$ using high-precision all-order method. 
The energies of the $9p_{1/2}$, $10p_{1/2}$, and $10p_{3/2}$
levels are predicted.  The results for atomic properties are compared with 
available theoretical and experimental data. 
The PNC amplitude for the $7s-6d_{3/2}$
transition is found to be $45.2\times10^{-11}i|e|a_0(-Q_W/N)$.
The dependence of the PNC amplitude on the choice of nuclear parameters is studied.
 This work provides a number of recommended values for yet unmeasured properties of Ra$^+$. 

\begin{acknowledgments}
 This work
was supported in part by National Science Foundation Grant  No.\
PHY-07-58088.
\end{acknowledgments}

\end{document}